\documentclass[conference]{IEEEtran}
\IEEEoverridecommandlockouts

\usepackage[T1]{fontenc}
\usepackage{xurl}
\usepackage{cite}
\usepackage{amsmath,amssymb,amsfonts}
\usepackage{algorithm}
\usepackage{algorithmic}
\usepackage{graphicx}
\usepackage{tikz}
\usepackage{textcomp}
\usepackage{xcolor}
\usepackage{pgfplots}
\usepackage{bm}
\usepackage{booktabs}

\usepackage{subcaption}
\usepackage{pgfplotstable}
\pgfplotsset{compat=1.18}
\usepackage{standalone}
\pgfplotsset{table/search path={figures}}
\usepackage{newtxmath}
\usepackage{balance}
\usepackage{siunitx}

\usepackage{placeins}
\usepgfplotslibrary{fillbetween}

\usetikzlibrary{arrows.meta,calc,decorations.markings,positioning,shapes.geometric}

\begin{document}
	
	\onecolumn
	\thispagestyle{empty}
	\vspace*{\fill}
	\begin{center}
		{\large This work has been submitted to the IEEE for possible publication.
			Copyright may be transferred without notice, after which this version may no
			longer be accessible.\par}
		\vspace{1.5em}
		Submitted to the 2026 IEEE Navigation Conference (NAVICON).
	\end{center}
	\vspace*{\fill}
	\clearpage
	\twocolumn

	\title{Achievable Accuracy and Cram\'er--Rao Bounds for
	SSB-Based LEO Positioning in NR NTN}
	
	\author{\IEEEauthorblockN{Rainer Bachl\IEEEauthorrefmark{1}, Tingting Lei\IEEEauthorrefmark{1}\IEEEauthorrefmark{2}, Muhammad Nabeel\IEEEauthorrefmark{1}}
		
		\IEEEauthorblockA{\IEEEauthorrefmark{1}{Huawei Heisenberg Research Center, Munich, Germany}}
		\IEEEauthorblockA{\IEEEauthorrefmark{2}{Institute of Astronomical and Physical Geodesy, Technical University of Munich, Munich, Germany}}
		\IEEEauthorblockA{\texttt{Email: \{rainer.bachl,tingting.lei1,muhammad.nabeel8\}@huawei.com
		}}
	}
	
	\definecolor{matlab1}{RGB}{0, 114, 189}      
	\definecolor{matlab2}{RGB}{217, 83, 25}      
	\definecolor{matlab3}{RGB}{237, 177, 32}     
	\definecolor{matlab4}{RGB}{126, 47, 142}     
	\definecolor{matlab5}{RGB}{119, 172, 48}     
	
	\maketitle
	
	\begin{abstract}
	Advanced Low Earth Orbit (LEO) satellite networks, such as Starlink's Mobile Satellite Service (MSS), will adopt the 5G New Radio (NR) Non-Terrestrial Network (NTN) standard.
	This enables the use of ubiquitous Synchronization Signal Blocks (SSBs) for opportunistic receiver positioning based on pseudorange and Doppler measurements.
	In this work, we characterize the estimation-theoretic limits of SSB-based positioning by deriving single-SSB Cramér--Rao lower bounds (CRLBs) for delay and carrier-frequency observables associated with pseudorange and Doppler.
	These bounds are obtained using the full SSB time-frequency energy distribution and extended into a multi-epoch, multi-satellite Fisher information framework that jointly bounds stationary-user position, clock bias, and clock drift.
	Each SSB contribution is weighted according to a range-dependent CRLB determined by the received SNR, so the estimator and the bound share a common noise model.
	In simulation, the resulting bound closely matches achievable performance, and a physically weighted least-squares estimator approaches the CRLB at realistic operating SNR.
	Using a Starlink-based constellation, we analyze the operating SNR experienced by a ground user and demonstrate sub-meter positioning accuracy.
	\end{abstract}
	
\section{Introduction}
\label{sec:intro}
Low Earth Orbit (LEO) mega-constellations are gaining significant attention as a complement to Global Navigation Satellite Systems (GNSS) in challenging environments~\cite{Stock2025}.
The lower orbit altitude of LEO satellites provides stronger received signal power and faster geometry variation than GNSS, resulting in substantially improved geometric diversity for navigation~\cite{Reid2018}.
Among existing systems, SpaceX's Starlink is the largest operational LEO constellation, with more than $10\,000$ satellites currently in orbit\footnote{\url{https://www.celestrak.org}}.
Recent filings with the Federal Communications Commission indicate that SpaceX plans to deploy a Mobile Satellite Service (MSS) compliant with the 3rd Generation Partnership Project (3GPP) New Radio (NR) Non-Terrestrial Network (NTN) standard for direct-to-cell services~\cite{SpaceXMSS}.
Unlike proprietary LEO systems, NR NTN transparently broadcasts accurate and up-to-date satellite ephemerides, greatly reducing the dominant source of error in opportunistic LEO-based positioning~\cite{Stock2025}.

Within the NR NTN framework, the dedicated Positioning Reference Signal (PRS) has been widely investigated for high-accuracy receiver positioning~\cite{Gonzalez2024,Dureppagari2025}.
However, PRS is an optional signal and is therefore not guaranteed to be available for positioning~\cite{Edjekouane2025}.
Moreover, allocating PRS resources reduces the overall network throughput~\cite{Dureppagari2025}.
These limitations motivate the use of the always-on Synchronization Signal Block (SSB), which is transmitted periodically and unconditionally by every LEO satellite for cell search and initial access.

Recent studies have demonstrated the positioning potential of SSB in LEO networks.
In~\cite{Zhu2024}, SSB is used to estimate timing advance for receiver positioning.
A comparison of PRS- and SSB-based time-of-arrival positioning is presented in~\cite{Edjekouane2025}, while~\cite{neinavaie2021acquisition} exploits SSB for Doppler tracking, an approach rooted in carrier-Doppler LEO navigation~\cite{Psiaki2021}.
In our recent work~\cite{Bachl2026}, we derived SSB-based pseudorange and Doppler measurement models for the downlink in LEO networks, and resolved the inherent $10$~ms per-satellite integer ambiguity using geometry alone.
While estimation-theoretic limits are well established for terrestrial positioning and unmanned aerial vehicle sensing~\cite{jopanya2025utilizing}, the corresponding bounds for opportunistic LEO positioning require dedicated analysis.
Therefore, in this work, we derive these bounds for opportunistic SSB positioning and benchmark the achievable accuracy against them.

Our main contributions are summarized as follows:

\begin{itemize}
	\item We derive Cram\'er--Rao lower bounds (CRLBs) for the Doppler and pseudorange observables from a single SSB, and propagate them into a multi-epoch, multi-satellite Fisher information matrix bounding the stationary-receiver position, clock bias, and clock drift (Sections~\ref{sec:measurements} and~\ref{sec:accuracy}). 		
	\item Building on the positioning framework of~\cite{Bachl2026}, we weight
	each SSB measurement by its own range-dependent variance, derived from
	the single-SSB CRLB evaluated at the SNR estimated from that reception.
	Estimator and bound thus share one physical noise model, in contrast to
	the flat measurement weighting used previously
	(Section~\ref{sec:wls}).
	\item On a Starlink-based constellation modeled on
	SpaceX's pending MSS filing, we benchmark the achieved positioning
	accuracy against the bound, characterize the operating SNR a ground
	user experiences, and identify the bias terms that would separate
	accuracy from the precision floor in deployment
	(Section~\ref{sec:results}).
	\end{itemize}
	
\section{Positioning Measurements and Bounds}
	\label{sec:measurements}
	
For opportunistic positioning, we focus solely on SSB.
According to 3GPP NR, SSB is broadcast with a short repetition period of $T_{\mathrm{SSB}}\in\{5,10,20,40,80,160\}$\,ms~\cite{TS38331}.
The two observables extracted from each SSB are a Doppler shift and a pseudorange, both relating to the unknown ground-user position and the receiver clock bias and drift.
This section explains the SSB structure, the measurements extracted from it, and their error bounds.

\subsection{SSB Signal Model}
\label{sec:ntn}

\begin{figure}[t]
	\centering
	\includegraphics[width=0.91\columnwidth]{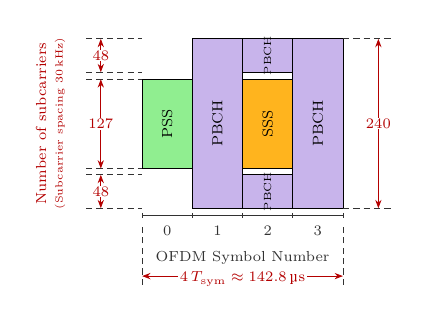}
	\caption{SSB time-frequency structure for the S-band NTN configuration with
		subcarrier spacing of $\SI{30}{\kilo\hertz}$.}
	\label{fig:ssb_structure}
\end{figure}

The SSB occupies four consecutive OFDM symbols and $240$ contiguous
subcarriers, forming a time-frequency grid of resource elements (REs)~\cite{TS38211}.
Figure~\ref{fig:ssb_structure} depicts the SSB structure for the S-band NTN
configuration with subcarrier spacing of $\Delta f_{\mathrm{SCS}}=\SI{30}{\kilo\hertz}$.
Both Primary Synchronization Signal (PSS) and Secondary Synchronization Signal (SSS) occupy the $127$ central subcarriers of the first and third symbols, respectively; the PBCH, together with its demodulation reference signal (DM-RS), fills the second and fourth symbols and additionally the remaining subcarriers around SSS.
With $\Delta f_{\mathrm{SCS}} = \SI{30}{\kilo\hertz}$, each symbol lasts
$T_{\mathrm{sym}}\approx\SI{35.7}{\micro\second}$, resulting in a single SSB duration of around $\SI{142.8}{\micro\second}$.
The receiver may assume equal energy per
resource element (EPRE) for the SSS, PBCH and PBCH DM-RS, and a PSS-to-SSS
EPRE ratio of either $0$ or $\SI{3}{\decibel}$~\cite{TS38213}.
We capture this single degree of freedom by the per-RE weight
$\beta_{\mathrm{PSS}}\in\{1,2\}$.

From the SSB, we target pseudorange and Doppler measurements.
The pseudorange requires the SSB transmit instant $T_{\mathrm{Tx}}$ in network time,
determined by the system frame number, the half-frame
indicator, and the SSB index, all of which are obtained by decoding
the PBCH and its DM-RS.
Without decoding the PBCH, the SSB yields a carrier-frequency measurement, i.e., Doppler, but no usable timing reference for ranging.
Once the PBCH is decoded, the entire SSB becomes deterministic at the receiver, and all $830$ REs can be exploited coherently for refined timing
and frequency estimation, instead of the initially known $398$ REs of the PSS, SSS, and DM-RS.

\subsection{Types of Measurements}

\subsubsection{Doppler}
\label{sec:doppler-meas}

The Doppler measurement for the $k$-th SSB of satellite $i$ is the
deviation of the received carrier frequency from its nominal value $f_c$,
\begin{equation}
	\Delta f_{i,k} = f^{D}_{i,k}(\mathbf r) + d\,f_c + \nu^{f}_{i,k},
	\qquad \nu^{f}_{i,k}\sim\mathcal N(0,\sigma_f^2),
	\label{eq:doppler}
\end{equation}
where $f^{D}_{i,k}$ is the geometric Doppler shift observed at the unknown ground-user position $\mathbf r$ due to satellite motion through the line-of-sight relation, and $d$
is the receiver clock drift (a fractional-frequency offset), shared across all satellites and epochs. The term $\nu^{f}_{i,k}$ is zero-mean Gaussian measurement noise of variance $\sigma_f^2$.
A line-of-sight velocity $v_{\mathrm{los}}$ shifts the carrier frequency by $-(f_c/c)\,v_{\mathrm{los}}$, with $c$ the speed of light, so \eqref{eq:doppler} is equivalently a range-rate observable with standard deviation $\sigma_{\dot\rho} = (c/f_c)\,\sigma_f$.
The full model is derived in~\cite{Bachl2026}.

\subsubsection{Pseudorange}
\label{sec:pseudorange-meas}

After resolving the per-satellite $10$~ms integer ambiguity by geometry
alone, as established in~\cite{Bachl2026}, the ambiguity-corrected
pseudorange for the $k$-th SSB of satellite $i$ is
\begin{equation}
	\mathrm{PR}_{i,k} = \rho_{i,k}(\mathbf r) + c\,b + c\,d\,t_{i,k}
	+ \nu^{\mathrm{pr}}_{i,k},
	\qquad \nu^{\mathrm{pr}}_{i,k}\sim\mathcal N(0,\sigma_\rho^2),
	\label{eq:pr}
\end{equation}
where $\rho_{i,k}$ is the geometric range, $b$ the receiver clock bias
(modulo $10$~ms after ambiguity resolution), $t_{i,k}$ the measurement
time relative to the batch origin, and $\nu^{\mathrm{pr}}_{i,k}$ the
corresponding pseudorange noise of variance $\sigma_\rho^2$. The noise terms
are mutually independent across satellites and epochs, so the network
covariance $\mathbf R$ is diagonal.

\subsection{Cram\'er--Rao Lower Bounds for the Measurements}
\label{sec:crlb-meas}

The variances $\sigma_f^2$ and $\sigma_\rho^2$ are not free parameters, and are bounded below by the information the SSB carries. We model one
received SSB in complex baseband as
\begin{equation}
	y[n] = \alpha\,e^{\,j2\pi f t_n}\,s[n;\tau] + w[n],
	\qquad w[n]\sim\mathcal{CN}(0,\sigma_w^2),
	\label{eq:sig}
\end{equation}
with $s[n;\tau]$ the known SSB waveform delayed by $\tau$, $f$ the residual
carrier-frequency offset, $\alpha$ an unknown complex gain, and $t_n=nT_s$
with $T_s$ the sampling interval.
For this deterministic signal in complex additive white Gaussian noise (AWGN), the Fisher information on the delay $\tau$ and frequency offset $f$ is set by the integrated (post-correlation) SNR of one SSB, $\gamma\triangleq|\alpha|^2 E_s/\sigma_w^2=\gamma_{\mathrm{RE}}\,B$~\cite{Kay1993}. Here $E_s=\sum_n|s[n]|^2$ is the signal energy, $\gamma_{\mathrm{RE}}$ the per-RE SNR after the FFT, and $B=\sum_{\mathrm{RE}}\beta_{\mathrm{RE}}$ the effective number of REs, where the per-RE EPRE weight $\beta_{\mathrm{RE}}$ equals $\beta_{\mathrm{PSS}}$ on the PSS resource elements and $1$ otherwise.
The carrier phase and amplitude are treated as nuisance parameters and eliminated by the Schur complement of the corresponding information block. This re-references the time
and frequency moments to the SSB energy centroid, so the relevant quantities are the central second moments. This yields the single-SSB bounds
\begin{equation}
	\sigma_f \ge \frac{1}{2\pi\,\sigma_t\sqrt{2\gamma}},
	\qquad
	\sigma_\rho = c\,\sigma_\tau
	\ge \frac{c}{2\pi\,W_{\mathrm{rms}}\sqrt{2\gamma}},
	\label{eq:crlb}
\end{equation}
with $\sigma_\tau$ the delay standard deviation. The energy-weighted RMS time spread and bandwidth are
$\sigma_t^2 = \langle t^2\rangle - \langle t\rangle^2$ and
$W_{\mathrm{rms}}^2 = \langle F^2\rangle - \langle F\rangle^2$.
The central moments use the energy-weighted averages $\langle t^p\rangle = E_s^{-1}\sum_n t_n^p\,|s[n]|^2$ and $\langle F^p\rangle = E_s^{-1}\sum_m F_m^p\,|S_m|^2$, with $t_n$ the
sample time of~\eqref{eq:sig} and $F_m$, $S_m$ the frequency and value of subcarrier $m$.
Equations~\eqref{eq:crlb} are the classical Gabor/Rife-Boorstyn bounds~\cite{RifeBoorstyn1974}, where frequency precision is set by the time aperture and delay precision by the bandwidth.
Since each SSB symbol is symmetric in frequency about the SSB center, the
time-frequency cross moment $\langle tF\rangle-\langle t\rangle\langle
F\rangle$ vanishes, the delay and Doppler sub-problems decouple, and
\eqref{eq:crlb} may be applied independently.

Evaluated over the full $830$-RE SSB map at $\Delta f_{\mathrm{SCS}}=\SI{30}{\kilo\hertz}$
($\beta_{\mathrm{PSS}}=1$), we obtain $\sigma_t\approx\SI{39}{\micro\second}$
and $W_{\mathrm{rms}}\approx\SI{1.96}{\mega\hertz}$, giving
$\sigma_f\approx\SI{101}{\hertz}$ and $\sigma_\rho\approx\SI{0.60}{\meter}$
at $\gamma_{\mathrm{RE}}=\SI{0}{\decibel}$, with the $\gamma^{-1/2}$
scaling of \eqref{eq:crlb}. The permitted $\SI{3}{\decibel}$ PSS boost
($\beta_{\mathrm{PSS}}=2$) raises $B$ from $830$ to $957$ and
tightens these to $\approx\SI{87}{\hertz}$ and $\approx\SI{0.58}{\meter}$.
Figure~\ref{fig:meas_crlb} plots both bounds against $\gamma_{\mathrm{RE}}$.

The per-RE SNR $\gamma_{\mathrm{RE}}$ at which these bounds are read is set by
the NTN link budget. For a flat power spectral density across the occupied
band, $\gamma_{\mathrm{RE}}$ equals the wideband SNR over the signal
bandwidth, so link-budget figures translate directly to the horizontal axis of
Fig.~\ref{fig:meas_crlb}. The 3GPP NTN link-budget methodology~\cite{TR38821}
and representative calibrated S-band results place the LEO downlink per-RE SNR
near $3.6$~dB at beam center~\cite{Khan2021}. We adopt this as the zenith
reference $\gamma_{\mathrm{RE}}^{\mathrm{zen}}=3.6$~dB, i.e.\ the closest, best-geometry case. As a satellite at approximately $326$~km altitude descends from zenith
to the $40^\circ$ elevation mask, the
free-space slant-range loss lowers the received SNR by up to $3.6$~dB, so the
simulated per-RE SNR spans roughly $0$ to $3.6$~dB, as discussed later in the results section. 
To confirm the single-SSB bounds hold more broadly, Fig.~\ref{fig:meas_crlb}
shades a wider operating range, $\gamma_{\mathrm{RE}}\approx-5$ to
$10$~dB, that also covers the beam-pattern roll-off and terminal variation of
a real deployment.

Per reception, the per-RE SNR obeys the free-space inverse-square law
$\gamma_{\mathrm{RE}}(\rho_{i,k})/\gamma_{\mathrm{RE}}^{\mathrm{zen}}=(\rho_{\mathrm{ref}}/\rho_{i,k})^2$ about the zenith reference range
$\rho_{\mathrm{ref}}$, and the integrated SNR follows as
$\gamma_{i,k}=\gamma_{\mathrm{RE}}(\rho_{i,k})\,B$. The per-SSB variances
$\sigma_{f,i,k}^2$ and $\sigma_{\rho,i,k}^2$ obtained from \eqref{eq:crlb}
at $\gamma_{i,k}$ are the weights used by the estimator and the network
bound in the next section.

\section{Positioning and the Network Bound}
\label{sec:accuracy}

We estimate the ground-user state from all SSB measurements and bound its
attainable precision by the network Fisher information.

\subsection{Range-Dependent Weighting and the Position Estimate}
\label{sec:wls}

The stationary-user state collects the position and clock terms $\mathbf x = [\mathbf r^\mathrm{T}\; b\; d]^\mathrm{T}\in\mathbb R^5$.
The noise-free measurement functions $h^{f}_{i,k}(\mathbf x) = \Delta f_{i,k} - \nu^{f}_{i,k}$ and $h^{\rho}_{i,k}(\mathbf x) = 	\mathrm{PR}_{i,k} - \nu^{\mathrm{pr}}_{i,k}$ are the deterministic parts of the Doppler~\eqref{eq:doppler} and pseudorange~\eqref{eq:pr} models.

Since each reception has its own integrated SNR $\gamma_{i,k}$, the per-SSB variances follow from \eqref{eq:crlb} as $\sigma^2_{\rho,i,k}=\sigma^2_\rho(\gamma_{i,k})$ and $\sigma^2_{f,i,k}=\sigma^2_f(\gamma_{i,k})$.
Without the receiver position, the slant range and resulting $\gamma_{i,k}$ are also unknown.
Therefore, the receiver estimates $\gamma_{i,k}$ directly from the received SSB through its $830$ REs after PBCH decoding, and uses this estimate in \eqref{eq:crlb} to form the weight.
Weighting by $1/\hat\sigma^2$ instead of $1/\sigma^2$ leaves the
least-squares estimate unbiased and inflates its variance only to second
order in the relative weight error, since the variance is stationary in the
weights at the optimum~\cite{Carroll1982}; the SNR-estimation error is
therefore negligible.

The state is estimated by weighted least squares as
\begin{equation}
	\hat{\mathbf x} = \arg\min_{\mathbf x}
	\sum_i \sum_{k(i)}
	\frac{\big(\mathrm{PR}_{i,k}-h^{\rho}_{i,k}(\mathbf x)\big)^2}
	{\sigma^2_{\rho,i,k}}
	+ \frac{\big(\Delta f_{i,k}-h^{f}_{i,k}(\mathbf x)\big)^2}
	{\sigma^2_{f,i,k}},
	\label{eq:wls}
\end{equation}
with $\mathrm{PR}_{i,k}$ the ambiguity-corrected pseudorange of \eqref{eq:pr},
the outer sum over the selected satellites, and the inner sum over the SSB
epochs of satellite $i$.
The batch spans a common grid of $N$ epochs at spacing $\Delta t$.
A selected satellite contributes the $M_i\le N$ epochs for which it stays above the elevation mask, with $M_i=N$ unless it sets mid-batch.
In contrast to~\cite{Bachl2026}, where a single pair $\sigma_\rho,\sigma_f$ weighted every measurement equally, here each residual carries its own range-dependent variance from the single-SSB CRLB.
As a result, high-SNR near-zenith receptions are weighted more strongly.
This nonlinear cost is minimized by the adaptive-weights damped Gauss-Newton
method with backtracking~\cite{Morichi2026}, initialized from a coarse position, following the framework of~\cite{Bachl2026}.

\begin{figure}[t]
	\centering
	\pgfplotsset{compat=1.18}
		\begin{tikzpicture}
		\def\xminval{-16}
		\def\xmaxval{16}
		\def\opmin{-5}
		\def\opmax{10}
		
		\begin{semilogyaxis}[
			width=0.92\linewidth,
			height=0.58\linewidth,
			xmin=\xminval, xmax=\xmaxval,
			ymin=1e-1, ymax=1e4,
			xlabel={per-RE SNR $\gamma_{\mathrm{RE}}$ (dB)},
			ylabel={CRLB},
			axis y line*=left,
			axis x line*=bottom,
			xtick={-15,-10,-5,0,5,10,15},
			ytickten={-1,0,1,2,3,4},
			set layers=standard,
			grid=both,
			major grid style={black!20},
			minor grid style={black!10},
			tick style={black},
			label style={font=\footnotesize},
			ticklabel style={font=\footnotesize},
			legend style={
				font=\footnotesize,
				draw=black,
				fill=white,
				at={(0.98,0.98)},
				anchor=north east
			},
			clip mode=individual
			]
			
			\addplot[
			draw=none,
			fill=black!10,
			on layer=axis background,
			forget plot
			] coordinates {(\opmin,1e-1) (\opmax,1e-1) (\opmax,1e4) (\opmin,1e4)} \closedcycle;
			
			\addplot[
			black,
			very thick,
			samples=200,
			domain=\xminval:\xmaxval
			]
			{1/(2*pi*38.755955548e-6*sqrt(2*830*10^(x/10)))};
			\addlegendentry{ $\sigma_{f,i,k}$ (Hz)}
			
			\addplot[
			black,
			very thick,
			dashed,
			samples=200,
			domain=\xminval:\xmaxval
			]
			{299792458/(2*pi*1.961584699e6*sqrt(2*830*10^(x/10)))};
			\addlegendentry{ $\sigma_{\rho,i,k}$ (m)}
			
			\node[font=\footnotesize, fill=white, inner sep=1pt]
			at (axis description cs:0.58,0.4) {expected operating range};
			
		\end{semilogyaxis}
	\end{tikzpicture}

		\caption{Single-SSB CRLB for the Doppler $\sigma_f$ and pseudorange $\sigma_\rho$ observable versus per-RE SNR $\gamma_{\mathrm{RE}}$, for the full $830$-RE SSB map
		($\beta_{\mathrm{PSS}}=1$, $\Delta f_{\mathrm{SCS}}=\SI{30}{\kilo\hertz}$,
		$f_c=\SI{2}{\giga\hertz}$). The shaded band marks the $-5$~dB to $10$~dB
		operating range.}
	\label{fig:meas_crlb}
\end{figure}
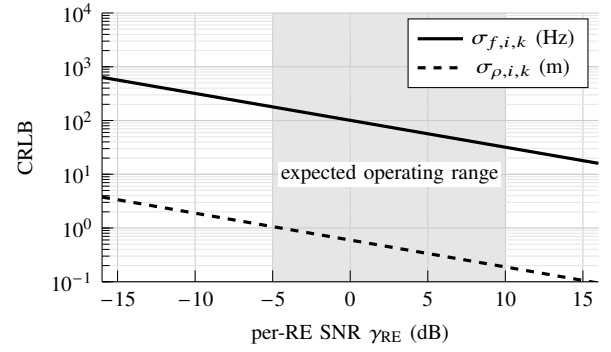
\vspace{-4mm}

\subsection{Cram\'er--Rao Lower Bounds for Positioning}
\label{sec:crlb-pos}

Linearizing $h^{\rho}_{i,k}$ and $h^{f}_{i,k}$ about the true state gives the Jacobian
rows $\mathbf h^{\rho}_{i,k}=\partial h^{\rho}_{i,k}/\partial\mathbf x$ and $\mathbf h^{f}_{i,k}=\partial h^{f}_{i,k}/\partial\mathbf x$,
\begin{equation}
	\mathbf h^{\rho}_{i,k} =
	\big[\,-\mathbf e_{i,k}^\mathrm{T},\; c,\; c\,t_{i,k}\,\big],
	\quad
	\mathbf h^{f}_{i,k} =
	\Big[\,\tfrac{f_c}{c\,\rho_{i,k}}\,\mathbf v_{\perp,i,k}^\mathrm{T},
	\; 0,\; f_c\,\Big],
	\label{eq:jac}
\end{equation}
where $\mathbf e_{i,k}$ is the line-of-sight unit vector and
$\mathbf v_{\perp,i,k}$ the satellite-velocity component orthogonal to it.
Here, the pseudorange senses position along
the line-of-sight direction $\mathbf e$, while the Doppler senses it through the
transverse velocity $\mathbf v_\perp/\rho$.
For LEO the transverse velocity is large, which renders the position observable even from short single-satellite arcs and separates the drift $d$ from the bias $b$.

Stacking the rows into $\mathbf H$ and the per-SSB variances into the
diagonal $\mathbf R$, the Fisher information matrix is
\begin{equation}
	\mathbf J = \sum_i \sum_{k(i)}
	\left[
	\frac{\mathbf h^{\rho\,\mathrm{T}}_{i,k}\mathbf h^{\rho}_{i,k}}
	{\sigma^2_{\rho,i,k}}
	+ \frac{\mathbf h^{f\,\mathrm{T}}_{i,k}\mathbf h^{f}_{i,k}}
	{\sigma^2_{f,i,k}}
	\right]
	= \mathbf H^\mathrm{T}\mathbf R^{-1}\mathbf H,
	\label{eq:fim_net}
\end{equation}
a $5\times5$ matrix, and the bound is $\mathrm{Cov}(\hat{\mathbf x})\succeq\mathbf J^{-1}$ in the positive-semidefinite sense.
The 3D position, clock-bias, and clock-drift standard deviations follow as
\begin{equation}
	\sigma_{\mathrm{pos}}=\sqrt{\operatorname{tr}[\mathbf J^{-1}]_{1:3,1:3}},
	\quad
	\sigma_b=\sqrt{[\mathbf J^{-1}]_{44}},
	\quad
	\sigma_d=\sqrt{[\mathbf J^{-1}]_{55}}.
	\label{eq:bounds}
\end{equation}
The clock states are carried in range units $c\,b$ and $c\,d$ during
inversion for numerical conditioning; the entries of $\mathbf J^{-1}$ are
scaled back by $c$ afterwards, so that $\sigma_b$ in \eqref{eq:bounds} is a
time (reported in ns) and $\sigma_d$ a dimensionless fractional-frequency
quantity (reported in ppb). The bound \eqref{eq:bounds} is evaluated for the Starlink-based constellation in the results section.

\subsection{From Precision to Accuracy}
\label{sec:bias-pos}

The CRLB bounds the estimator variance, but any bias increases the mean squared error as $\mathrm{MSE}=\mathrm{bias}^2+\mathrm{variance}$.
In the scenario considered here, every bias term is zero or negligible
by construction. In particular, the measurement noise is zero-mean, the clock bias and
drift are estimated states rather than biases, and the $10$~ms integer
ambiguity has been resolved by geometry~\cite{Bachl2026}. The
ionospheric and tropospheric delays can be precompensated to a sub-noise
residual~\cite{SanzSubirana2013}. The System
Information Block 19 (SIB19) ephemeris is usually only seconds old~\cite{TS38331}, so its propagation error is negligible.
Finally, the $O(\sigma^2)$ bias of the nonlinear
least-squares inversion itself~\cite{Box1971} evaluates to approximately $10^{-8}$~m
for this geometry, far below $\sigma_{\mathrm{pos}}$. The variance bound
\eqref{eq:bounds} therefore doubles as the accuracy floor in this setting,
and the Monte Carlo accuracy of the next section approaches it.

\section{Results and Discussion}
\label{sec:results}

This section describes the simulation setup and then evaluates the proposed approach.

\subsection{Simulation Setup}
\label{sec:sim_setup}

We evaluate both the bound and the estimator on the Starlink MSS
constellation of Table~\ref{tab:constellation}: $12\,000$ satellites in four Walker-delta shells between $326$~km and $332$~km, with inclinations and plane counts consistent with SpaceX's
pending MSS filing~\cite{SpaceXMSS}.
The satellites are propagated with SGP4 to the SSB transmit instants only. 
The constellation generation, orbit propagation, integer-ambiguity resolution and weighted-least-squares solver follow the framework of~\cite{Bachl2026}.

A satellite is considered visible above $40^\circ$ elevation.
With the ground user located in Munich, more than eight satellites are visible at all times.
Of the visible satellites, up to eight are selected for the batch by largest positive Doppler, i.e., the rising satellites.
This guarantees the longest remaining visibility and lets the selected satellites climb toward closest approach over the batch.
Beams are Earth-fixed with distinct per-satellite SSB transmit timings.
When a selected satellite sets, its beam is taken over by the next rising satellite.


Unlike a fixed per-link error model, the pseudorange and Doppler standard deviations are set per SSB by the single-SSB CRLB, evaluated at the integrated SNR of that reception. The per-reception SNR follows the free-space slant-range law, anchored at a reference value $\gamma_{\mathrm{RE}}^{\mathrm{zen}} = 3.6$~dB~\cite{Khan2021}, defined when a satellite in the lowest shell ($326$~km) is at zenith.
The resulting $\sigma_{\rho,i,k}$ and $\sigma_{f,i,k}$ weight both the estimator~\eqref{eq:wls} and the bound~\eqref{eq:fim_net}.

Each reported positioning result is the RMS error over $4000$ Monte Carlo runs.
A run draws independent realizations of four quantities:
\begin{itemize}
	\item The frequency-offset noise $\nu^{f}_{i,k}$ and pseudorange noise $\nu^{\mathrm{pr}}_{i,k}$ at the per-SSB standard deviations.
	\item The receiver clock bias $b$, uniform on $[0,10\,\mathrm{ms})$ to exercise the full ambiguity interval	of~\cite{Bachl2026}.
	\item The clock drift $d$, uniform with $|d|\le0.1$~ppm.
	\item The initial solver position, Gaussian with $100$~km standard deviation. 
\end{itemize} 
A fixed seed set is reused across configurations so that any two are compared on identical realizations.
Except for the instantaneous-CRLB trajectory, which sweeps the batch start across the observation window, the batch start time is held fixed so that the geometry is identical across configurations and the observed spread reflects the error sources alone.

The channel is open-sky line-of-sight with AWGN. Tropospheric and ionospheric delays are not modeled, since both can be compensated to below the measurement noise. At $f_c=2$~GHz, the ionospheric delay is suppressed as $1/f_c^2$ and remains at the meter level for typical electron content. The tropospheric slant delay is frequency-independent and larger, approximately $4$~m at $40^\circ$ elevation~\cite{SanzSubirana2013}. Both exceed $\sigma_\rho$, but are compensated upfront from a coarse a priori position, leaving a decimeter-level residual below $\sigma_\rho$.
Other parameters are listed in Table~\ref{tab:simulation_param}.

\begin{table}[t]
	\centering
	\caption{Constellation based on Starlink MSS. The Walker-delta configuration is given as inclination${:}\,t/p/f$ (total satellites\,/\,planes\,/\,phasing).}
	\setlength{\tabcolsep}{2pt}
	\begin{tabular*}{\columnwidth}{@{\hspace{2pt}\extracolsep{\fill}}c c c@{\hspace{2pt}} r@{\extracolsep{0pt}\,:\,}l@{\extracolsep{\fill}}  c@{\hspace{2pt}}}
		\toprule
		\textbf{Orbital} & \textbf{Altitude} &\textbf{RAAN of 1st}& \multicolumn{2}{c}{\textbf{Walker-delta}} &  \textbf{Percentage} \\
		\textbf{Shell}   & \textbf{(km)}  & \textbf{orbital plane}    & \multicolumn{2}{c}{\textbf{configuration}}   & \textbf{of Sats.} \\
		\midrule
		1 & 332 & $0^\circ$ & $43^\circ$   & $4080/68/16$  & 34\% \\
		2 & 330 & $90^\circ$   & $53^\circ$ &  $5760/96/23$  & 48\% \\
		3 & 328 & $180^\circ$   & $69^\circ$ &  $840/28/6$   &  7\% \\
		4 & 326 &  $270^\circ$ & $96.87^\circ$ & $1320/22/5$  & 11\% \\
		\midrule
		\multicolumn{3}{r}{\textbf{Total number of Satellites:}} & \multicolumn{2}{c}{\textbf{12000}} & \textbf{100\%} \\
		\bottomrule
	\end{tabular*}
	\label{tab:constellation}
\end{table}

\begin{table}[t]
	\centering
	\caption{Simulation Parameters}
	\begin{tabular}{ll}
		\toprule
		\textbf{Parameter} & \textbf{Value} \\
		\midrule
		Carrier frequency & $f_c = 2$ GHz \\
		Subcarrier spacing & 30 kHz \\
		SSB period & 160 ms \\
		Beams per satellite & 128 \\
		Max. sats providing beam coverage & 8 \\
		Satellite selection criterion & Max. positive Doppler \\
		Min. elevation for visibility & 40\textdegree \\
		Ground user location & Munich (48.14\textdegree N, 11.58\textdegree E) \\
		Max. clock bias (uniform dist.) & $10$ ms \\
		Max. clock drift (uniform dist.)& $0.1$ ppm \\
		Std. dev. init. pos. error (normal dist.) & 100~km \\
		Monte Carlo runs per pos. error result &  4000 \\ 		
		\bottomrule
	\end{tabular}
	\label{tab:simulation_param}
\end{table}

\subsection{Simulation Results}
\label{sec:sim_results}

\begin{figure}[t]
	\centering
	\includegraphics[width=\columnwidth]{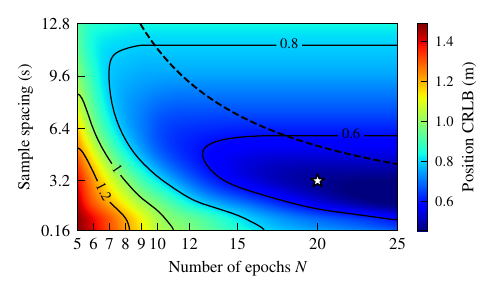}
	\caption{Position CRLB over the number of epochs $N$ and sample spacing
		$\Delta t$. The dashed line marks the maximum satellite-visibility period at $330$~km altitude, corresponding to
		$T_{\mathrm{batch}}=$ \mbox{$(N-1)\Delta t$ = } $101.3$~s. The marker is the selected $(N^\star,\Delta t^\star)$.}
	\label{fig:crlb_contour}
\end{figure}

Figure~\ref{fig:crlb_contour} shows the 3D position CRLB over the number of
epochs $N$ and the spacing $\Delta t$, the latter taken as multiples of the
$160$~ms SSB period, evaluated in closed form from~\eqref{eq:bounds} at the
$3.6$~dB zenith SNR reference. Two effects are visible. Under the
independent-epoch (diagonal-$\mathbf R$) model the information is a monotone
sum over epochs, so increasing $N$ always tightens the bound.
The spacing improves the bound not by noise averaging but through geometric diversity.
At fixed $N$, a wider spacing samples a larger line-of-sight rotation and velocity change, improving the conditioning of $\mathbf H^\mathrm{T}\mathbf R^{-1}\mathbf H$.
This benefit saturates and then reverses once the batch spans a change in the satellite set or visibility, so the bound has an interior optimum in $\Delta t$ rather than improving monotonically.
We adopt $N^\star= 20$ epochs at $\Delta t^\star= 3.2$~s
($T_{\mathrm{batch}}=60.8$~s) for the remainder.
Since the CRLB is inversely proportional to $\sqrt{\gamma}$,
shifting the reference SNR scales the entire surface of Fig.~\ref{fig:crlb_contour} uniformly.
The relative performance and the selected $(N^\star,\Delta t^\star)$ therefore remain valid at any reference SNR.

\begin{figure}[tbp]
	\centering
	\includegraphics{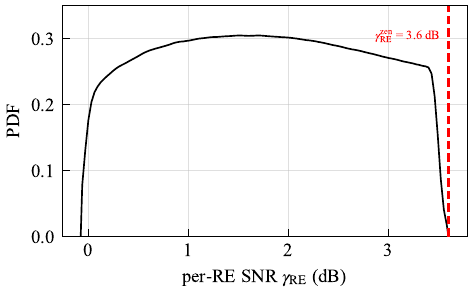}
	\caption{Distribution of per-RE SNR 
		for the selected satellites, at the 
		reference level of $3.6$~dB for a satellite at zenith.}
	\label{fig:snr_distribution}
	\vspace{-3mm}
\end{figure}

Figure~\ref{fig:snr_distribution} shows the per-RE SNR received from the selected satellites.
Since the selection favors rising satellites near the mask, the distribution spreads downward from the $3.6$~dB zenith reference and stays within the operating range of Fig.~\ref{fig:meas_crlb}, keeping every reception in the CRLB-valid regime.

\begin{figure*}[t]
	\centering
	{\includegraphics[width=0.62\columnwidth]{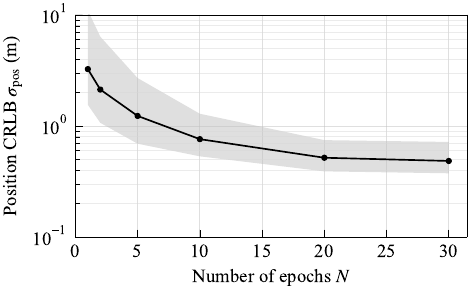}}\qquad
	{\includegraphics[width=0.62\columnwidth]{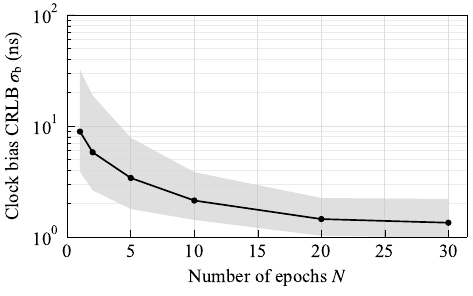}}\qquad
	{\includegraphics[width=0.62\columnwidth]{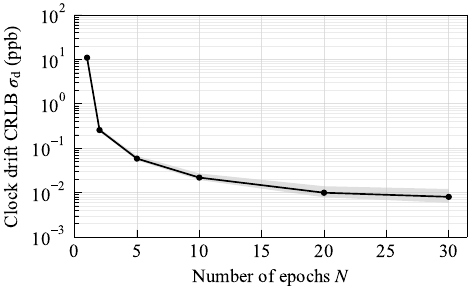}}
	\caption{Mean of CRLB~\eqref{eq:bounds} (with area between 1st and 99th percentile shaded) for position, clock bias and clock drift versus
		number of epochs $N$ at $\Delta t^\star = 3.2$~s spacing at $1000$ batch start times with different satellite geometries; the percentile spread reflects geometry variation.}
	\label{fig:crlb_vs_N}
	\vspace{-5mm}
\end{figure*}

With $\Delta t^\star$ fixed, we now show in Fig.~\ref{fig:crlb_vs_N} the position, clock-bias, and clock-drift CRLB as $N$ grows, evaluated at $1000$ batch start times placed across the constellation's passes.
The shaded area between the 1st and 99th percentiles reflects the geometry variation across start times rather than measurement noise.
The clock drift becomes observable only for $N\ge2$ and its CRLB decreases rapidly with increasing $N$. Position and clock-bias CRLB tighten smoothly with increasing $N$. There is virtually no further positioning improvement when increasing $N$ from 20 to 30, which is consistent with Fig.~\ref{fig:crlb_contour} and motivates the choice of $N^\star =$ 20.
At $N^\star$ the bound for position is $0.52$~m, for clock
bias $1.45$~ns, and for drift $0.01$~ppb.

\begin{figure}[t]
	\centering
	\includegraphics{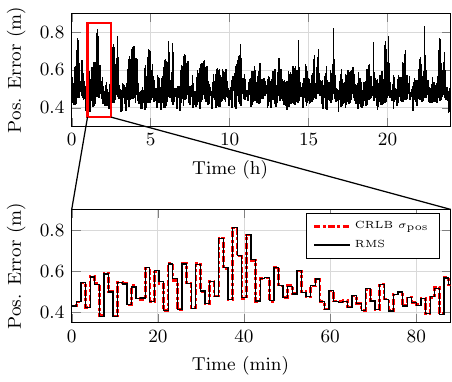}
	\caption{Instantaneous position CRLB with satellites selected at the start of every batch window and the matching Monte Carlo RMS error over 4000 runs that replay the same
		satellite geometry and randomize the error sources. The zoomed time range in the lower subplot corresponds approximately to one satellite revolution.}
	\label{fig:traj_positioning}
	\vspace{-3mm}
\end{figure}

Finally, Fig.~\ref{fig:traj_positioning} replays the time-varying constellation to plot the resulting instantaneous position CRLB.
At each batch start time, it forms a fresh $(N^\star,\Delta t^\star)$ batch
from the visible satellites and inverts that batch's Fisher matrix.
The matching curve is the RMS error over $4000$ Monte Carlo runs that replay
the same geometry at each time and randomize only the frequency,
pseudorange, clock-bias, clock-drift and initial-position errors.
The bound fluctuates between $0.37$ and $0.83$~m.
Moreover, the Monte Carlo RMS error rides on the instantaneous bound across the entire window, so the estimator attains the precision floor over the full range of geometries the constellation presents, not only at a single operating point.

\vspace{-10mm}

\subsection{Discussion}
The achievable 3D positioning accuracy is $0.52$~m under idealized zero-mean and bias-free noise.
This is better than the meter-scale accuracy achieved with the flat-weighted model in~\cite{Bachl2026} because the per-SSB CRLB assigns the high-SNR receptions their true sub-meter pseudorange precision rather than a single conservative value.
The split between the observables is asymmetric.
At the operating SNR, the pseudorange CRLB is sub-meter while the Doppler CRLB is at the $100$~Hz level; therefore, the main contribution to the high positioning accuracy comes from the pseudorange measurements, consistent with the measurement sensitivity reported in~\cite{Bachl2026}.
As discussed earlier in Section~\ref{sec:bias-pos}, the negligible systematic error
makes this precision floor the accuracy floor in simulation.
In deployment, two bias terms reopen the gap: the residual atmospheric delay after coarse
compensation (decimeter-level at S-band, growing at low elevation) and the
SIB19 ephemeris quantization (of order $0.65$~m per satellite from the
$1.3$~m position step of~\cite{TS38331}, a bias over
arcs shorter than the ephemeris-validity window).
Both are slowly varying and add in quadrature to the RMS error, consistent with the tens-of-meters, bias-dominated budgets reported for SSB/PRS NTN positioning in~\cite{Edjekouane2025}.

\section{Conclusion}
\label{sec:conclusion}

We derived the estimation-theoretic limits of opportunistic SSB-based LEO
positioning in NR NTN. Starting from the full SSB time-frequency specification, single-SSB CRLBs for the Doppler and pseudorange
observables were obtained and propagated into a multi-epoch,
multi-satellite Fisher matrix bounding the user position, clock bias, and
clock drift, with each measurement weighted by a range-dependent variance. Systematic error terms are negligible by construction in
the considered scenario, which makes the precision floor coincide with the
accuracy floor. Evaluated on a constellation based on Starlink MSS over Munich, the
weighted least-squares estimator attains a 3D accuracy in the sub-meter range, tracking the instantaneous bound across the full range of
geometries the constellation presents. 
The bounds established here
define the limits of what SSB-based LEO positioning can achieve without any modification
to the NR NTN standard.

\balance

\bibliographystyle{IEEEtran}
\bibliography{conference_paper2_copy_for_arXiv}

\end{document}